\documentclass[]{spie}
\usepackage[T1]{fontenc}
\usepackage[utf8]{inputenc}
\usepackage{mathptmx}
\usepackage{amsmath,amssymb}
\usepackage{graphicx}
\usepackage{booktabs}
\usepackage[protrusion=false]{microtype}
\usepackage{url}

\usepackage[font=small,labelsep=period,justification=justified,singlelinecheck=false,
            margin={0.3in,0pt}]{caption}

\makeatletter
\def\subsectfont{\raggedright\normalsize\bf}
\def\acknowledgements{\section*{ACKNOWLEDGEMENTS}}
\def\thebibliography#1{\section*{REFERENCES\@mkboth{REFERENCES}{REFERENCES}}
  \list{[\arabic{enumi}]}
  {\settowidth\labelwidth{[99]}\leftmargin\labelwidth
   \advance\leftmargin\labelsep \usecounter{enumi}
   \itemsep 0pt \parsep 0pt \topsep 0pt}
  \def\newblock{\hskip .11em plus .33em minus .07em}
  \sloppy\clubpenalty4000\widowpenalty4000 \sfcode`\.=1000\relax}
\makeatother

\title{Analyzing image encoder choices and graph homophily in GCN frameworks for breast ultrasound classification}

\author{Sabahattin~Mert~Daloglu\supit{*a}, Ceren~Coskun\supit{a}, Harvey~Castro\supit{a},
Soner~Hacihaliloglu\supit{a}, and Ilker~Hacihaliloglu\supit{a,b}
\skiplinehalf
\supit{a}PONS Incorporated, Newark, NJ, USA; \\
\supit{b}Department of Radiology, Department of Medicine, University of British Columbia, \\
Vancouver, BC, Canada
}

\authorinfo{*md@ponstech.co (corresponding author). Other author e-mail addresses: cc@ponstech.co (C.C.), harvery@ponstech.co (H.C.), soner@ponstech.co (S.H.), ilker.hacihaliloglu@ubc.ca (I.H.)}

\begin{document}
\maketitle

\begin{abstract}
Breast ultrasound is widely used for screening, yet automated analysis remains challenging due to speckle noise, acquisition variability, and weak separation of benign and malignant cases in standard ultrasound imaging. Graph convolutional networks (GCNs) have recently emerged as a promising approach by leveraging relationships among similar patient samples. However, it remains unclear how the choice of image encoder influences graph construction and downstream classification performance. In this work, we systematically evaluate five image encoders spanning convolutional and transformer-based architectures for GCN-based breast ultrasound classification. Image embeddings are used to construct cosine similarity k-nearest-neighbor graphs, which are classified using a single-layer GCN with a linear classification head. Across three patient-wise cross-validation folds, higher-capacity encoders consistently improve graph homophily and downstream classification performance, yielding gains in accuracy, AUC, sensitivity, specificity, and F1-score. Moreover, test-set graph homophily exhibits a strong linear correlation with classification accuracy (R-squared = 0.875), with higher-capacity encoders consistently occupying the high-homophily, high-accuracy region, suggesting that encoder-driven improvements in graph structure are a key mechanism underlying the observed performance gains. On 798 images from two external sources, the encoder ordering was preserved, but accuracy was 0.13--0.21 lower than in cross-validation and the homophily--accuracy association was weaker (R-squared = 0.759). These findings establish encoder selection as a critical factor in graph-based breast ultrasound classification and identify graph homophily as a key indicator linking representation quality to downstream classification performance.
\end{abstract}


\section{INTRODUCTION}
\label{sec:intro}

Breast cancer remains one of the leading causes of cancer-related mortality among women worldwide. Early detection is critical for improving patient outcomes, and ultrasound is widely used as a complementary imaging modality to mammography due to its cost-effectiveness, portability, and suitability for dense breast tissue. Despite these advantages, automated analysis of breast ultrasound remains challenging due to speckle noise, operator-dependent acquisition variability, and subtle visual differences between benign and malignant lesions.\cite{luo2024deeplearningbreast}

Recent advances in deep learning have significantly improved performance in medical image analysis, with convolutional neural networks (CNNs)\cite{he2016resnet} and transformer-based models such as the vision transformer (ViT)\cite{dosovitskiy2020vit} demonstrating strong representation learning capabilities. While several approaches aggregate information across multiple images or views from the same patient, they generally do not explicitly model relationships between different patients that may provide complementary diagnostic information. Graph convolutional networks (GCNs) address this limitation by representing each patient sample (here, each ultrasound scan) as a node in a similarity graph, enabling predictions to leverage information from related samples.\cite{DALOGLU2026100691,montaha2024gcn} Such relational modeling is particularly attractive for breast ultrasound, where high intra-class variability and subtle inter-class differences often make individual sample classification difficult.

A key factor influencing GCN performance is graph homophily, defined as the tendency of connected nodes to share similar labels. In the broader GCN literature, many studies have analyzed the homophily principle and proposed methods to better handle heterophilous graphs that are common in real-world settings.\cite{DBLP:journals/corr/abs-2106-06134,luanhomophily2024} Standard GCN-style models are fundamentally tied to the homophily assumption because they propagate and aggregate features over local graph neighborhoods. However, in breast ultrasound classification, the role of image encoder choice in shaping graph structure, specifically its impact on homophily and downstream classification performance, remains underexplored. In particular, it is unclear how different representation learning paradigms, ranging from lightweight CNNs to large-scale self-supervised ViTs such as DINO (self-distillation with no labels) and I-JEPA (image-based joint-embedding predictive architecture), influence the quality of similarity graphs constructed for GCN-based classification. Rather than proposing a new heterophily-robust GCN variant, we investigate how to increase graph homophily through better encoder representations and thereby improve GCN performance.

In previous work,\cite{DALOGLU2026100691} GCN-based breast ultrasound classification was investigated using a CNN backbone encoder. Here, we extend that analysis by isolating encoder design under a single B-mode input representation and systematically evaluating five image encoders, including a lightweight baseline CNN,\cite{DALOGLU2026100691} ResNet-18,\cite{he2016resnet,shia2021resnetbus} and three transformer-based self-supervised models: a masked autoencoder (MAE) ViT,\cite{he2021mae} DINO ViT-S/8,\cite{caron2021dino} and I-JEPA ViT.\cite{assran2023ijepa} Features extracted from each encoder are used to construct cosine similarity k-nearest-neighbor graphs, which are subsequently processed using a single-layer GCN with a linear classification head. The main contributions of this work are as follows:
\begin{itemize}
\setlength{\itemsep}{1pt}\setlength{\parskip}{0pt}
    \item \textbf{Comprehensive backbone comparison:} We present a systematic evaluation of five distinct image encoder families within a unified GCN framework for breast ultrasound classification, enabling controlled analysis of representation quality.
    \item \textbf{Consistent performance trend with stronger encoders:} We provide empirical evidence that higher-capacity encoders tend to shift GCN performance toward a more favorable operating regime, with higher accuracy, area under the receiver operating characteristic curve (AUC), sensitivity, specificity, and F1-score across folds.
    \item \textbf{Homophily-driven performance analysis:} We introduce a homophily-centric perspective that explicitly links encoder choice to graph structure and downstream GCN performance, and we report a strong linear association between graph homophily and classification accuracy.
\end{itemize}

\section{DATASET AND METHODS}
\label{sec:methods}

\subsection{Breast ultrasound dataset}
We combine scans from eight publicly available breast-ultrasound datasets\cite{hovertrans_tmi_2023,bus_uclm_mendeley_2025,us3m_tdfnet_2024,breast_lesions_usg_2024,ardakani2023breast_lesion_db,hamyoon2022bi_rads_morphometry,homayoun2022radiomics_multicenter,MH2017breast_ultrasound_lesion_detection,bus_bra_zenodo_2023,ALDHABYANI2020104863} (Table~\ref{tab:dataset_composition}). To form a binary classification dataset, we discard all normal scans and retain the benign/malignant labels provided by the original sources. This yields a final set of 6619 usable benign/malignant scans.

\begin{table}[!htb]
\caption{Composition of the eight publicly available datasets used. Normal scans are discarded (counts shown for transparency). After removing normals, we obtain 6619 benign/malignant scans.}
\label{tab:dataset_composition}
\begin{center}
\small
\setlength{\tabcolsep}{6pt}
\begin{tabular}{lcccc}
\toprule
Dataset & Benign & Malignant & Normal (discarded) & Total \\
\midrule
HoVer-Trans\cite{hovertrans_tmi_2023} & 886 & 1519 & 0 & 2405 \\
BUS-UCLM\cite{bus_uclm_mendeley_2025} & 174 & 90 & 419 & 683 \\
US3M\cite{us3m_tdfnet_2024} & 453 & 328 & 0 & 781 \\
Breast-Lesions-USG\cite{breast_lesions_usg_2024} & 154 & 98 & 4 & 256 \\
QAMEBI\cite{ardakani2023breast_lesion_db,hamyoon2022bi_rads_morphometry,homayoun2022radiomics_multicenter} & 109 & 123 & 0 & 232 \\
Breast-BUS\cite{MH2017breast_ultrasound_lesion_detection} & 109 & 54 & 0 & 163 \\
BUS-BRA\cite{bus_bra_zenodo_2023} & 1268 & 607 & 0 & 1875 \\
BUSI\cite{ALDHABYANI2020104863} & 437 & 210 & 0 & 647 \\
\midrule
Total & 3590 & 3029 & 423 & 7042 \\
\bottomrule
\end{tabular}
\end{center}
\end{table}

We partition the data into three patient-wise folds. For each fold, we target a 10\% test set with roughly balanced per-sub-dataset test allocation ($\approx$80 scans per source); smaller sources and integer rounding introduce slight fold-wise variability while preserving 6619 scans per fold. Table~\ref{tab:fold_distribution} summarizes the exact percentages and scan counts.

\begin{table}[!htb]
\caption{Fold-wise data distribution. Percentages and scan counts (in parentheses) are shown for the training, validation, and test splits in each fold.}
\label{tab:fold_distribution}
\begin{center}
\small
\setlength{\tabcolsep}{10pt}
\begin{tabular}{lccc}
\toprule
Fold & Train (\%) & Validation (\%) & Test (\%) \\
\midrule
1 & 80.3 (5316) & 10.0 (663) & 9.7 (640) \\
2 & 80.3 (5316) & 10.0 (663) & 9.7 (640) \\
3 & 81.4 (5391) & 10.2 (673) & 8.4 (555) \\
\bottomrule
\end{tabular}
\end{center}
\end{table}

\subsection{Backbone encoders}
\begin{figure}[t]
\begin{center}
\includegraphics[width=0.88\textwidth]{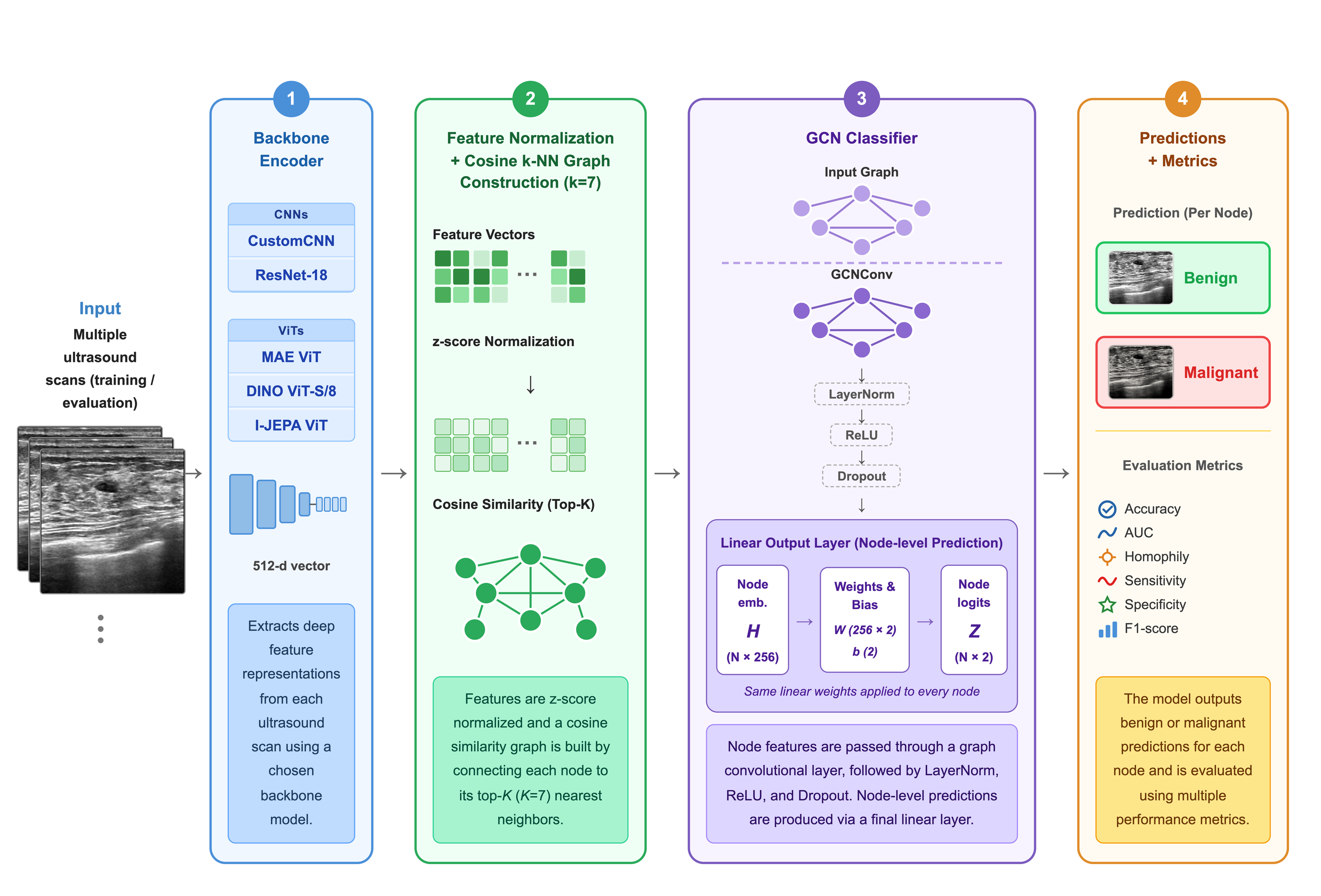}
\end{center}
\caption{Overview of the GCN pipeline. Ultrasound scans are encoded by one backbone (CustomCNN, ResNet-18, MAE ViT, DINO ViT-S/8, or I-JEPA ViT), z-score normalized, converted to a cosine top-$K$ graph ($K=7$), and processed by a GCN with one graph convolutional layer followed by a linear output layer to produce benign/malignant predictions and evaluation metrics (accuracy, AUC, homophily, sensitivity, specificity, and F1-score).}
\label{fig:pipeline_bmode}
\end{figure}

We evaluate five image encoder architectures spanning a range of representational capacity, including both convolutional and transformer-based models (Fig.~\ref{fig:pipeline_bmode}). Throughout this paper, the term \emph{backbone} refers to the image encoder that extracts representations prior to graph construction. For ResNet-18, DINO ViT-S/8 and I-JEPA ViT, we adopt publicly available architectures and initialize from publicly available pretrained weights. MAE ViT uses the same architecture and the same ultrasound-domain pretrained weights as in our previous work.\cite{DALOGLU2026100691} CustomCNN is trained from scratch. All backbones take $224\times224$ inputs and produce 512-dimensional feature vectors from their penultimate layers.
\begin{itemize}
\setlength{\itemsep}{1pt}\setlength{\parskip}{0pt}
    \item \textbf{CustomCNN} is a lightweight convolutional network composed of three convolutional blocks (32, 64, and 128 channels) with $3\times3$ kernels, batch normalization, Leaky ReLU activations, and $2\times2$ max-pooling, followed by a fully connected layer that maps $128\times28\times28$ activations to 512 features with dropout (0.5), and a final 2-class classification layer. This architecture is the baseline CNN encoder used in previous work.\cite{DALOGLU2026100691}
    \item \textbf{ResNet-18}\cite{he2016resnet,shia2021resnetbus} follows the standard four-stage residual design with 64, 128, 256, and 512 channels; we extract a global representation from the final stage and project it to 512 dimensions via a linear layer before a 2-class head.
    \item \textbf{DINO ViT-S/8} is a self-supervised vision transformer encoder,\cite{caron2021dino} with 12 transformer blocks (embedding dimension 384, 6 attention heads); features from selected layers are linearly projected (384$\rightarrow$512) and aggregated into a 512-dimensional vector.
    \item \textbf{MAE ViT} uses a ViT-Base encoder pre-trained with masked autoencoding;\cite{he2021mae} the weights come from an ultrasound-domain checkpoint (Sec.~\ref{sec:training}). We similarly extract multi-layer representations (embedding dimension 768) and project them to 512 dimensions.
    \item \textbf{I-JEPA ViT}\cite{assran2023ijepa} is based on an I-JEPA-pretrained transformer encoder, from which we extract features from representative transformer layers and project the model-specific embedding dimension to 512 via a linear layer.
\end{itemize}

\subsection{Graph construction and graph convolutional network}
For each fold, we build separate cosine-similarity graphs for the training, validation, and test splits. In each graph, nodes correspond to individual scans and edges connect samples with similar representations. Before graph construction, we apply z-score normalization to all node features. Let $\mathbf{z}_i \in \mathbb{R}^{d}$ denote the normalized feature vector of node $i$. Cosine similarity between nodes $i$ and $j$ is
\begin{equation}
\label{eq:cos}
    s_{ij} = \frac{\mathbf{z}_i^\top \mathbf{z}_j}{\lVert \mathbf{z}_i\rVert_2 \, \lVert \mathbf{z}_j\rVert_2}.
\end{equation}
Self-similarities are removed and, for each node, undirected edges are retained to the top-$K$ neighbors ($K=7$). We set $K=7$ after initial hyperparameter sweeps, as this neighborhood size provided a balanced graph sparsity level: small enough to preserve class-consistent local structure, yet large enough to maintain robust connectivity for reliable information diffusion during GCN updates.

Following Kipf and Welling,\cite{kipf2016gcn} with adjacency $A$, self-looped adjacency $\tilde{A}=A+I$, degree matrix $\tilde{D}$, and normalized adjacency $\hat{A}$,
\begin{equation}
\label{eq:prop}
\hat{A}=\tilde{D}^{-1/2}\tilde{A}\tilde{D}^{-1/2}.
\end{equation}
The classifier is a single-layer GCN followed by a linear output layer. A graph convolution maps node features from input dimension $d$ to 256 hidden channels using normalized neighborhood aggregation:
\begin{align}
H^{(0)} &= \hat{A}XW^{(0)} + \mathbf{b}^{(0)}, \\
H^{(1)} &= \mathrm{Dropout}\!\left(\mathrm{ReLU}\!\left(\mathrm{LayerNorm}(H^{(0)})\right)\right), \\
Z &= H^{(1)}W^{(1)} + \mathbf{b}^{(1)},
\end{align}
where $X$ is the node-feature matrix, $W^{(0)}$ and $W^{(1)}$ are trainable weight matrices, $\mathbf{b}^{(0)}$ and $\mathbf{b}^{(1)}$ are the convolution and output biases, $H^{(0)}$ and $H^{(1)}$ are hidden representations, $Z$ contains the class logits, and feature dropout uses $p{=}0.5$. The model is trained with cross-entropy loss; final predictions are obtained by argmax over the two logits (benign/malignant). This design is a practical implementation of the normalized GCN formulation of Kipf and Welling.\cite{kipf2016gcn}

Test-graph homophily, i.e., the fraction of test-graph edges that connect nodes sharing the same ground-truth label, is obtained using
\begin{equation}
\label{eq:homo}
    h = \frac{1}{|E|}\sum_{(i,j)\in E}\mathbb{I}[y_i = y_j],
\end{equation}
where $E$ is the set of edges in the test graph, $y_i$ is the ground-truth label of node $i$, and $\mathbb{I}[\cdot]$ is the indicator function.

\subsection{Training details and backbone pretraining}
\label{sec:training}
All images are resized to $224\times224$. Backbone training uses Adam (learning rate $10^{-4}$, weight decay $10^{-4}$, batch size 32) for up to 50 epochs with early stopping patience 10. The GCN is trained with Adam (learning rate $10^{-3}$, weight decay $10^{-4}$), hidden width 256, for up to 125 epochs with early stopping patience 50. For both backbone and GCN stages, model selection uses validation-accuracy early stopping. Every backbone--GCN configuration was trained in two independent runs of the identical pipeline; for each fold, all reported metrics (including test-graph homophily) are the average of the two runs. ROC curves pool the test predictions of all folds and both runs.

Backbone initialization and fine-tuning protocol are as follows. The lightweight CNN baseline (CustomCNN) is trained from scratch. ResNet-18 is initialized from ImageNet-pretrained weights.\cite{he2016resnet} DINO ViT-S/8 is initialized from publicly released self-supervised DINO weights.\cite{caron2021dino} MAE ViT-Base/16 is initialized from MAE pretraining\cite{he2021mae} and, in our runs, loaded from an ultrasound-domain MAE checkpoint pretrained on more than 230,000 deidentified ultrasound images from multiple public datasets.\cite{vandenheuvel2018fetal,meyer2025ultrasam,sappia2024acouslic,leclerc2019echonet,chen2021uscl,wang2018multifeature,mohabir2020knee,li2022breastlesion,ouyang2020video,reddy2023pediatriclvef} This is the same checkpoint as in our previous work.\cite{DALOGLU2026100691} I-JEPA ViT is initialized from publicly released JEPA pretraining.\cite{assran2023ijepa}

For transfer learning, we freeze the backbone at initialization and fine-tune the task head plus the last backbone block/group (last residual stage for ResNet, last transformer block for ViT-family encoders). The projection head and classifier layers remain trainable in all settings. Figure~\ref{fig:pipeline_bmode} summarizes the end-to-end experimental pipeline used for all backbones.

\subsection{External evaluation}
To examine whether the findings extend beyond the eight training sources, the trained models were evaluated, without retraining or tuning, on images from the public BUS-CoT collection\cite{buscot2025,buscot_figshare}\relax, which compiles breast ultrasound images with histopathology labels from several sources. We removed records originating from datasets that are part of our training collection (BUS-BRA, BUSI and Breast-BUS, which BUS-CoT labels UDIAT), kept only images with a benign or malignant label, drew one image per patient, and removed images that were near-duplicates of a training scan according to a 64-bit perceptual hash. Of the remaining sources we used TFUSL and TUHD; the others consist of images collected from web pages and published figures, with figure panels, overlaid text and widely varying image sizes. This gave 798 images from 798 patients (384 benign, 414 malignant; majority-class accuracy 0.519), using the cropped ultrasound field supplied by BUS-CoT. All 30 trained models (five encoders, three folds, two runs) were applied: features were extracted with the same pre-processing, a cosine top-$K$ graph ($K=7$) was built over all external images after z-score normalization, and each fold's GCN head classified the nodes. As for the internal results, values are averaged over the two runs of a fold and reported as mean$\pm$std over the three folds; 95\% confidence intervals come from 1,000 bootstrap resamples of the images.

\section{RESULTS}
\label{sec:results}

\subsection{Overall metrics across backbones}
Table~\ref{tab:bmode_acc_wide} summarizes 3-fold averaged results across all reported metrics (each fold averaged over two training runs). The backbone ordering is consistent in our experiments: the lightweight CNN baseline is lowest, ResNet-18 and MAE ViT are intermediate, and DINO ViT-S/8 and I-JEPA ViT occupy the highest-performance regime. This pattern appears jointly in accuracy, AUC, homophily, sensitivity, specificity, and F1-score, indicating that encoder quality is associated with both discriminative performance and graph structure. DINO ViT-S/8 is the strongest overall model, with the highest mean value in every reported metric. These results suggest that stronger encoder representations may produce cleaner neighborhood structure for message passing and, in turn, better downstream GCN classification.

\begin{table}[!htb]
\caption{GCN test metrics (mean$\pm$std over 3 folds; each fold value is the average of two independent training runs). Bold indicates the highest value per column.}
\label{tab:bmode_acc_wide}
\begin{center}
\small
\setlength{\tabcolsep}{4.5pt}
\begin{tabular}{lcccccc}
\toprule
GCN backbone & Accuracy & Homophily & AUC & Sensitivity & Specificity & F1-score \\
\midrule
CustomCNN    & 0.752$\pm$0.014 & 0.685$\pm$0.008 & 0.824$\pm$0.027 & 0.752$\pm$0.027 & 0.752$\pm$0.006 & 0.713$\pm$0.014 \\
MAE ViT      & 0.783$\pm$0.027 & 0.729$\pm$0.016 & 0.851$\pm$0.016 & 0.781$\pm$0.038 & 0.786$\pm$0.034 & 0.746$\pm$0.028 \\
ResNet-18    & 0.794$\pm$0.008 & 0.759$\pm$0.020 & 0.852$\pm$0.011 & 0.803$\pm$0.009 & 0.787$\pm$0.016 & 0.761$\pm$0.008 \\
I-JEPA ViT   & 0.817$\pm$0.009 & 0.780$\pm$0.011 & 0.886$\pm$0.002 & 0.828$\pm$0.011 & 0.811$\pm$0.019 & 0.788$\pm$0.019 \\
DINO ViT-S/8 & \textbf{0.846$\pm$0.011} & \textbf{0.811$\pm$0.010} & \textbf{0.905$\pm$0.011} & \textbf{0.858$\pm$0.019} & \textbf{0.839$\pm$0.029} & \textbf{0.820$\pm$0.013} \\
\bottomrule
\end{tabular}
\end{center}
\end{table}

\subsection{ROC curves by backbone}
\begin{figure}[t]
\begin{center}
\includegraphics[width=0.78\textwidth]{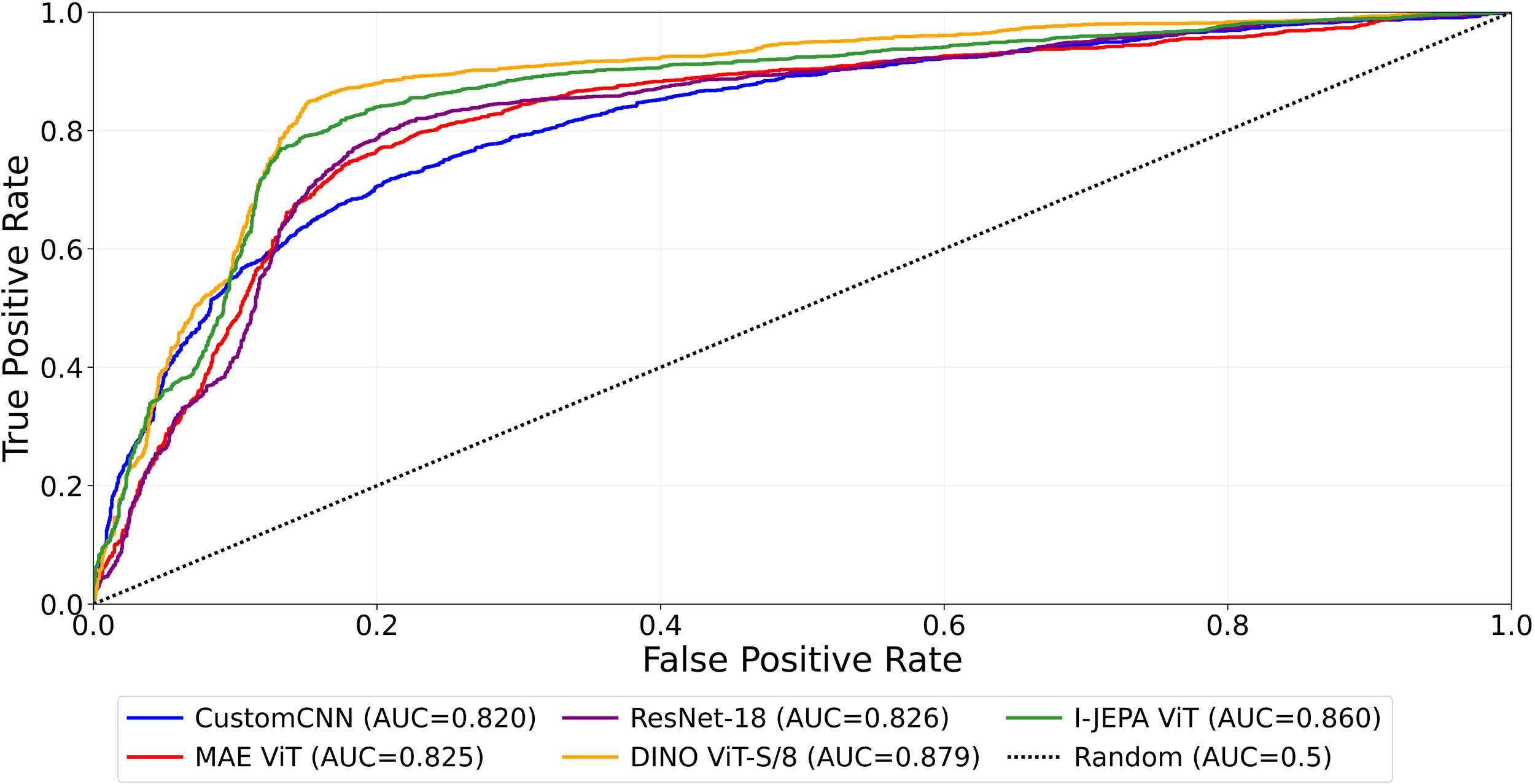}
\end{center}
\caption{Combined ROC plot (all backbones on one axis). For each backbone, predictions from all test folds and both training runs were pooled to compute a single ROC curve; legend AUC values correspond to the pooled ROC AUC.}
\label{fig:roc_bmode_acc_combined}
\end{figure}

Figure~\ref{fig:roc_bmode_acc_combined} shows the combined receiver operating characteristic (ROC) visualization across all backbones. The ROC shapes are consistent with Table~\ref{tab:bmode_acc_wide}: DINO ViT-S/8 and I-JEPA ViT dominate the upper-left region, with DINO ViT-S/8 achieving the strongest overall separation and I-JEPA ViT closely following. ResNet-18 and MAE ViT occupy a middle regime, while the lightweight CNN baseline has the lowest pooled AUC, although its curve lies above those of ResNet-18 and MAE ViT at very low false positive rates (the three pooled AUCs are close, 0.820--0.826). This ranking is consistent with the view that higher-capacity self-supervised representations can improve the quality of node features used for graph construction. The AUC values shown in the ROC legend are \emph{pooled} ROC AUC values (all fold and run predictions pooled before ROC computation), whereas Table~\ref{tab:bmode_acc_wide} reports fold-averaged AUC (mean$\pm$std). Although the absolute values differ due to the aggregation method, both views show the same backbone dependence trend in magnitude ordering.

\subsection{Accuracy versus homophily fit}
\begin{figure}[!tb]
\begin{center}
\includegraphics[width=0.86\textwidth]{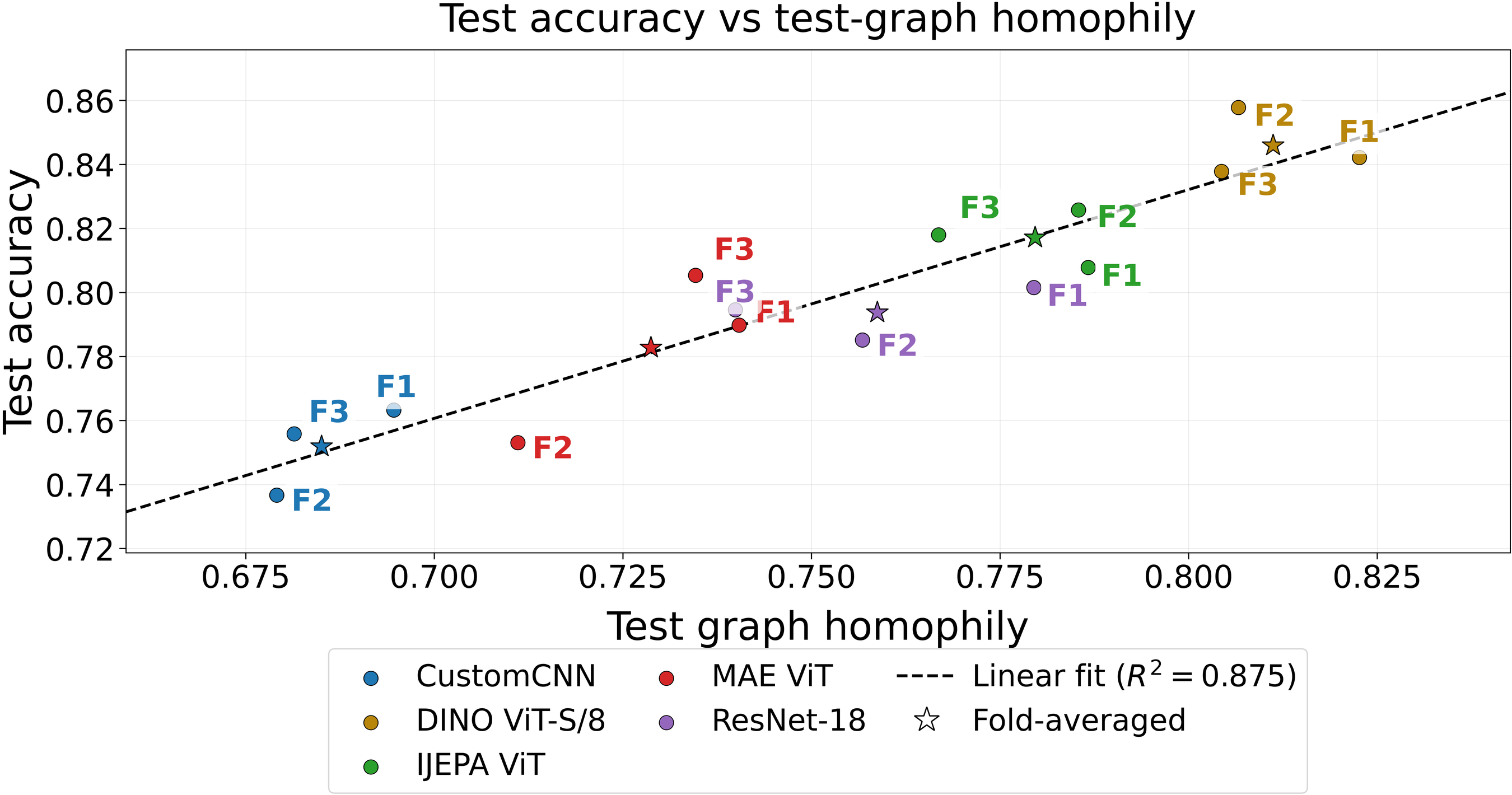}
\end{center}
\caption{Test accuracy vs.\ test-graph homophily for GCN models using fold-wise points (15 total: 3 folds per backbone, each the mean of two training runs), with one additional per-backbone average marker for visualization. The linear fit ($R^2=0.875$) is computed only from the fold-wise points.}
\label{fig:bmode_allfold_homophily_vs_acc}
\end{figure}

Figure~\ref{fig:bmode_allfold_homophily_vs_acc} shows the fold-wise relationship between test accuracy and test-graph homophily (15 points: 3 folds per backbone, each the mean of two training runs), plus one per-backbone average marker. The fit shows a strong positive association between homophily and test accuracy ($R^2=0.875$), with higher-capacity encoders consistently occupying the high-homophily, high-accuracy region. The regression fit is computed only from the 15 fold-wise points; per-backbone average markers are shown only as visual summaries and are not included in the fitting procedure.

\subsection{External evaluation}
\begin{figure}[!tb]
\begin{center}
\includegraphics[width=0.78\textwidth]{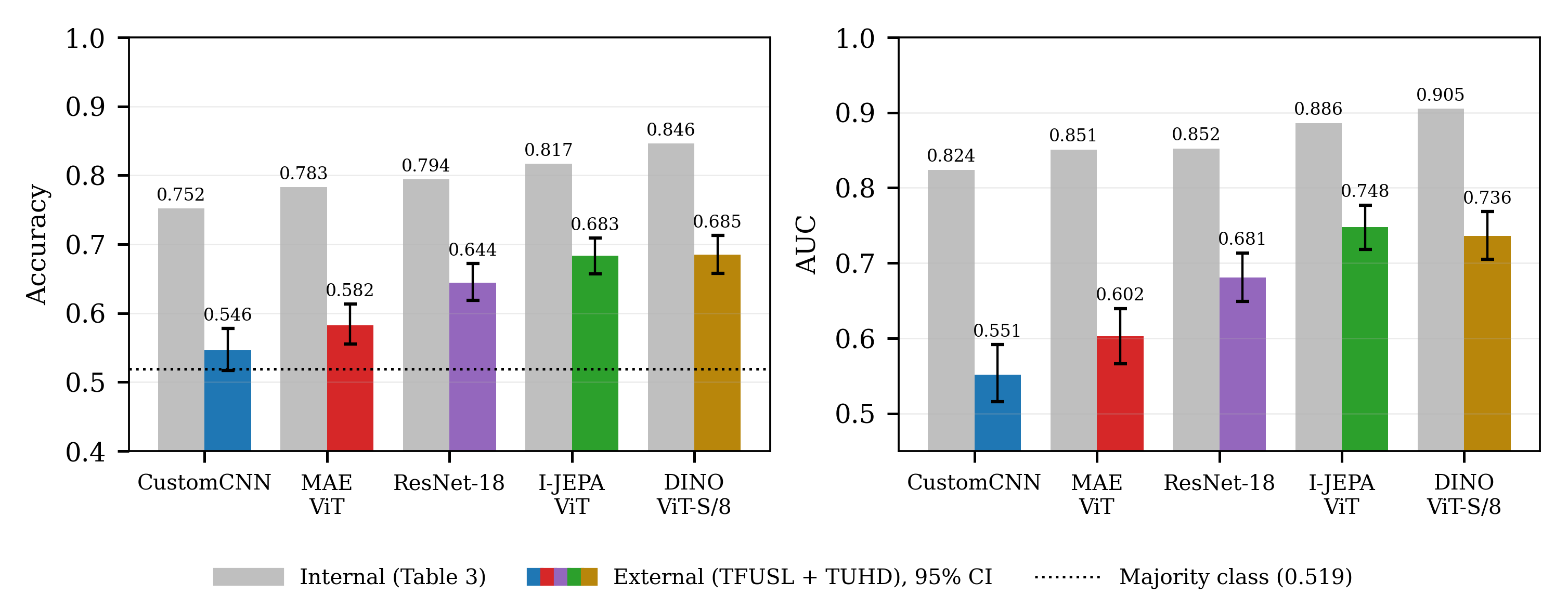}
\end{center}
\caption{Accuracy and AUC of each encoder in cross-validation (grey, Table~\ref{tab:bmode_acc_wide}) and on the external images (colour, with 95\% bootstrap interval). The dotted line is the majority-class accuracy (0.519).}
\label{fig:external}
\end{figure}

\begin{figure}[!tb]
\begin{center}
\includegraphics[width=0.45\textwidth]{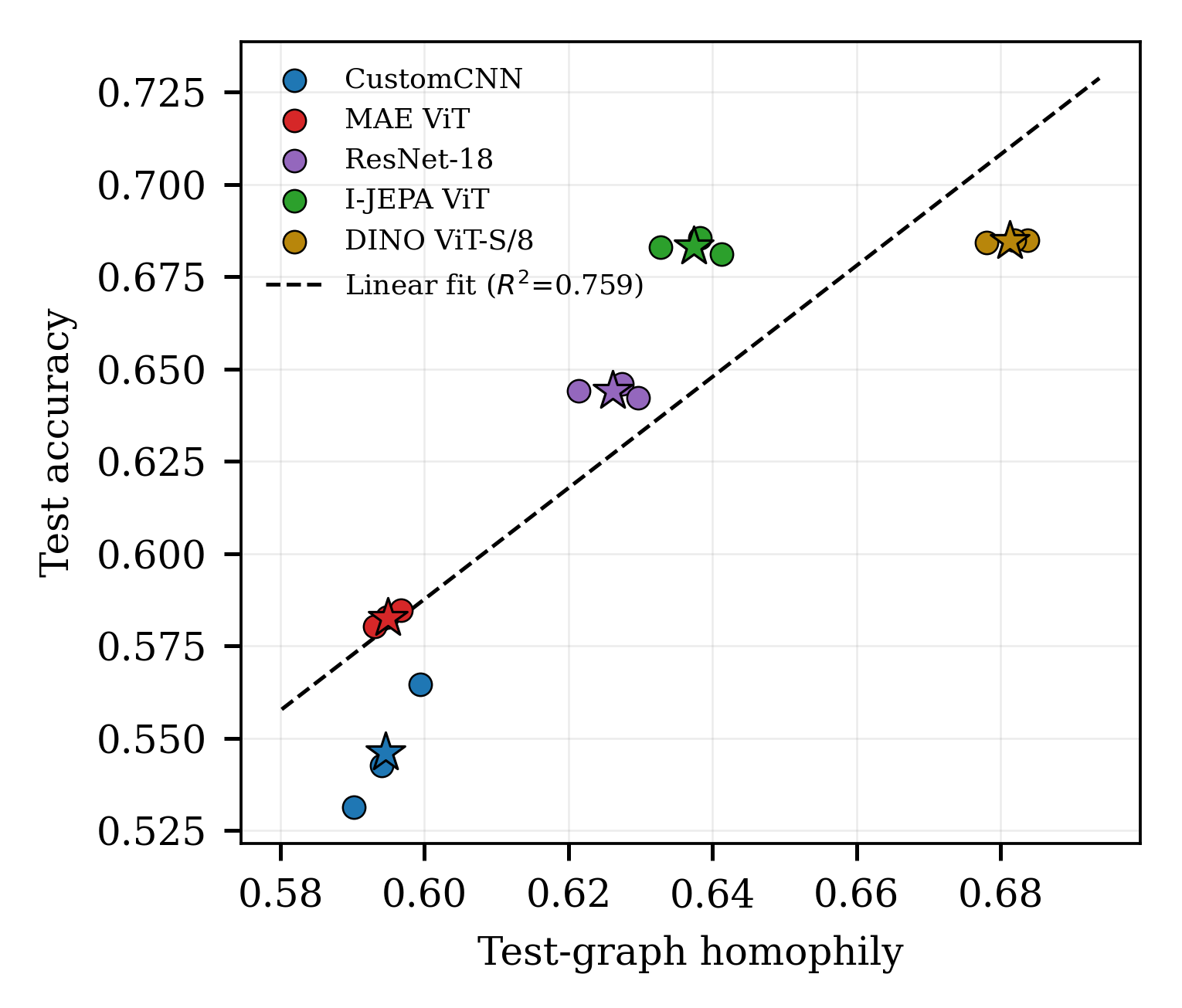}\hfill\includegraphics[width=0.45\textwidth]{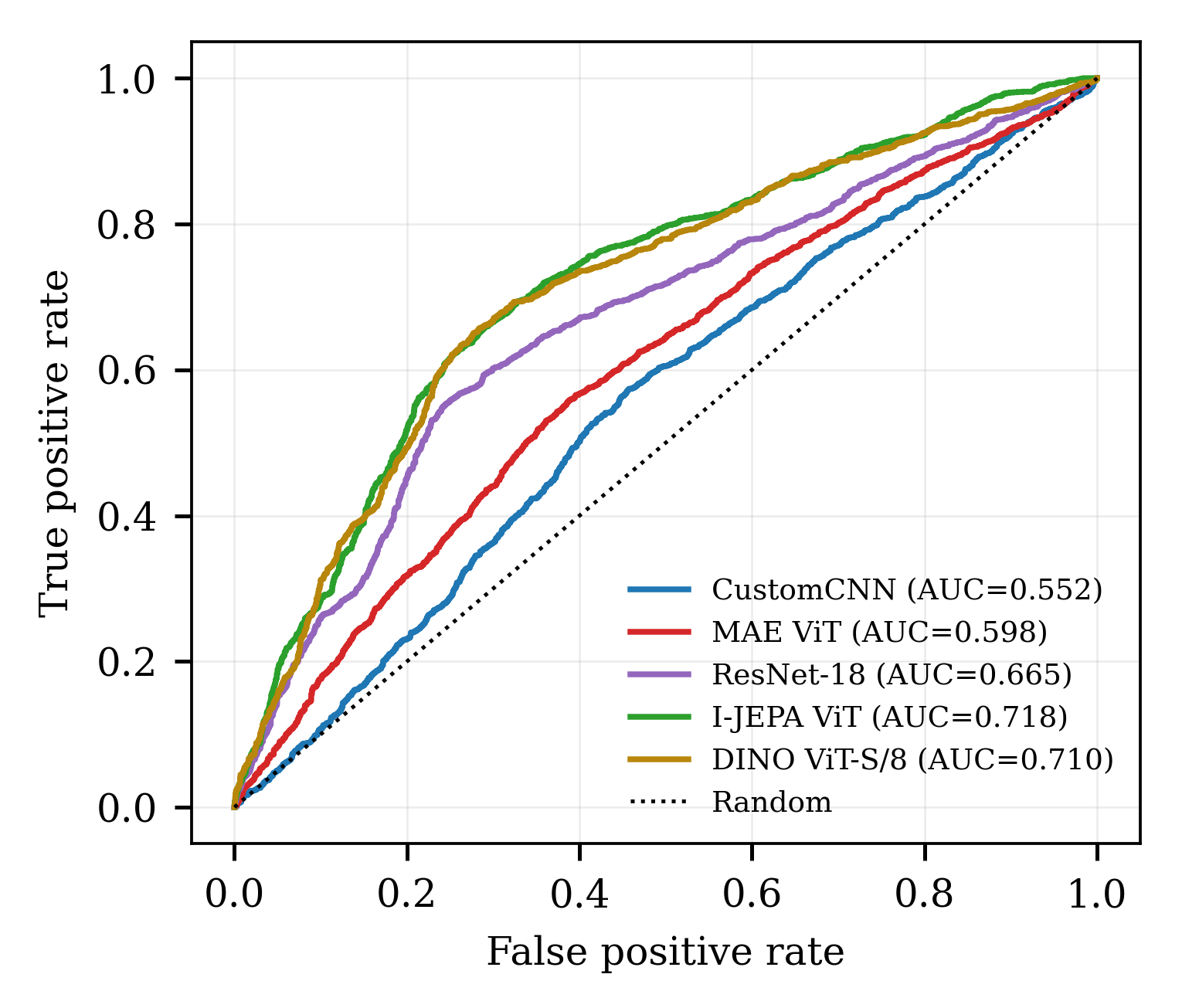}
\end{center}
\caption{Left: external test accuracy versus test-graph homophily (15 fold-wise points, each the mean of two training runs; stars are per-backbone means; the dashed line is the linear fit, $R^2=0.759$, computed from the fold-wise points). Right: ROC curves of each backbone on the external images, with the predictions of all folds and both runs pooled; legend values are pooled AUCs.}
\label{fig:external_hom}
\end{figure}

Table~\ref{tab:external} and Fig.~\ref{fig:external} report the results on the external images. Accuracy was lower than in cross-validation for every encoder (by 0.13--0.21; AUC by 0.14--0.27), and the lightweight CNN was close to the majority-class accuracy of 0.519. The ordering of the encoders by accuracy was the same as in cross-validation (CustomCNN $<$ MAE ViT $<$ ResNet-18 $<$ I-JEPA ViT $<$ DINO ViT-S/8). DINO ViT-S/8 (0.685, 95\% CI 0.658--0.713) and I-JEPA ViT (0.683, CI 0.657--0.709) were the best encoders; their accuracies differ by only 0.002 and their confidence intervals overlap almost completely, so neither can be called more accurate. Test-graph homophily was lower than in cross-validation but followed the same order as accuracy (CustomCNN and MAE ViT tied), and remained positively associated with accuracy across the 15 fold-level points (Fig.~\ref{fig:external_hom}, left; $R^2=0.759$, compared with $R^2=0.875$ in cross-validation). The ROC curves (Fig.~\ref{fig:external_hom}, right) show the same ordering of encoders, except that I-JEPA ViT had a slightly higher pooled AUC than DINO ViT-S/8 (0.718 versus 0.710).

\begin{table}[!htb]
\caption{GCN test metrics on the external images (798 BUS-CoT images from TFUSL and TUHD; mean$\pm$std over 3 folds, each fold the average of two training runs). Accuracy is given with its 95\% bootstrap interval; the last column is external minus cross-validation accuracy (Table~\ref{tab:bmode_acc_wide}). The majority-class accuracy is 0.519. Bold indicates the highest value per column.}
\label{tab:external}
\begin{center}
\small
\setlength{\tabcolsep}{4.5pt}
\begin{tabular}{lcccc}
\toprule
GCN backbone & Accuracy (95\% CI) & Homophily & AUC & $\Delta$ accuracy \\
\midrule
CustomCNN    & 0.546$\pm$0.017 [0.517, 0.578] & 0.595$\pm$0.005 & 0.551$\pm$0.011 & $-$0.206 \\
MAE ViT      & 0.582$\pm$0.002 [0.555, 0.613] & 0.595$\pm$0.002 & 0.602$\pm$0.012 & $-$0.201 \\
ResNet-18    & 0.644$\pm$0.002 [0.618, 0.672] & 0.626$\pm$0.004 & 0.681$\pm$0.015 & $-$0.150 \\
I-JEPA ViT   & 0.683$\pm$0.002 [0.657, 0.709] & 0.637$\pm$0.004 & \textbf{0.748$\pm$0.003} & $-$0.134 \\
DINO ViT-S/8 & \textbf{0.685$\pm$0.000 [0.658, 0.713]} & \textbf{0.681$\pm$0.003} & 0.736$\pm$0.006 & $-$0.161 \\
\bottomrule
\end{tabular}
\end{center}
\end{table}

\section{DISCUSSION}
\label{sec:discussion}

Under a fixed GCN architecture, stronger encoders improved both test-graph homophily and classification metrics (Table~\ref{tab:bmode_acc_wide}), suggesting that encoder choice shapes the similarity graph and downstream message passing. DINO ViT-S/8 achieved the best overall performance, followed closely by I-JEPA ViT, demonstrating the effectiveness of self-supervised vision transformer representations for GCN-based breast ultrasound classification. Figure~\ref{fig:bmode_allfold_homophily_vs_acc} further shows a strong association between homophily and test accuracy ($R^2=0.875$), supporting homophily as a useful cross-encoder predictor of GCN performance. Although GCN performance depends on multiple factors, these findings support the hypothesis that constructing graphs with more class-consistent neighborhoods enables more effective message passing and improved classification.

To our knowledge, this is the first study to systematically link image-encoder choice to graph homophily in a GCN-based breast ultrasound classification framework. Compared with prior CustomCNN-focused GCN work,\cite{DALOGLU2026100691} this study extends the analysis to multiple backbone architectures and provides evidence consistent with encoder-driven improvements in graph structure being a key mechanism underlying GCN performance gains in this application. It also suggests homophily as a practical, encoder-agnostic indicator for anticipating downstream GCN performance, and thus a practical criterion for encoder selection in future graph-based ultrasound classification systems.

The external evaluation supports the ordering of the encoders and a positive association between homophily and accuracy, but not the absolute cross-validation accuracies, which were 0.13--0.21 higher than on images from outside the training sources. Possible reasons are that the cross-validation test folds share sources and acquisition styles with the training folds, and that the external images carry burned-in annotations and come from heterogeneous scanners. The association between homophily and accuracy was weaker externally ($R^2=0.759$ versus $0.875$), and I-JEPA ViT matched DINO ViT-S/8 in accuracy despite lower homophily (0.637 versus 0.681), so homophily does not fully determine accuracy.

\textbf{Limitations.} Several limitations should be noted. First, the relationship between homophily and accuracy is based on 15 fold-wise points (three folds for each of five encoders), and the experiments do not isolate graph structure from other differences among encoders, such as pretraining data and objective, model size, patch size, and the number of backbone parameters that are fine-tuned. For example, MAE ViT, the only encoder initialized from an ultrasound-domain checkpoint, ranked below the DINO and I-JEPA encoders; separating the effects of pretraining domain, objective, and architecture is left for future work. Second, results are reported for a single GCN architecture, a fixed neighborhood size ($K=7$), and B-mode input only; alternative graph constructions and heterophily-robust graph neural networks were not explored. Future work should include controlled pretraining ablations and broader graph-construction studies.

\section{CONCLUSION}
\label{sec:conclusion}

We investigated encoder choice in GCN classification for breast ultrasound and showed that higher-capacity encoders improve both classification performance and graph homophily. Compared with previous CustomCNN-focused GCN work,\cite{DALOGLU2026100691} this extends the analysis to multiple backbone architectures, demonstrating that encoder choice substantially influences graph construction and downstream classification performance, and identifying graph homophily as a key indicator linking representation quality to downstream classification performance. On external images from outside the training sources, the encoder ordering and a positive homophily--accuracy association were preserved, at substantially lower absolute accuracy.

\acknowledgements
Research reported in this work was supported by the National Institute on Minority Health and Health Disparities of the National Institutes of Health under award number 1R43MD020006-01A1. The content is solely the responsibility of the authors and does not necessarily represent the official views of the National Institutes of Health. In accordance with SPIE policy, the authors disclose that Claude (Anthropic), a large language model, was used to assist with drafting, wording refinement, grammar correction, and formatting into the SPIE template. The study design, experiments, and analyses are the authors' own work. The authors reviewed and verified all content, results, and references, and take full responsibility for the manuscript.

\end{document}